\documentclass[aps,prb,twocolumn,showpacs,superscriptaddress,amsmath,amssymb]{revtex4}
\usepackage{xcolor}
\usepackage{graphicx}
\usepackage{amsmath}
\usepackage{amssymb}
\usepackage{epsfig}
\usepackage{epstopdf}
\usepackage{wasysym}
\usepackage{bbm}
\usepackage[colorlinks,citecolor=blue,linkcolor=blue]{hyperref}
\usepackage{multirow}

\begin{document}

\title{Stability, Electronic and Magnetic properties of magnetically doped
topological insulators Bi$_{2}$Se$_{3}$, Bi$_{2}$Te$_{3}$ and
Sb$_{2}$Te$_{3}$}

\author{Jian-Min Zhang}

\affiliation{Beijing National Laboratory for Condensed Matter Physics and Institute
of Physics,\linebreak{}
 Chinese Academy of Sciences, Beijing 100190, China}

\author{Wenmei Ming}

\affiliation{Department of Materials Science and Engineering, University of Utah,
Salt Lake City 84112, USA}

\author{Zhigao Huang}

\affiliation{Department of Physics, Fujian Normal University, Fuzhou 350007, China}

\author{Gui-Bin Liu}

\affiliation{School of Physics, Beijing Institute of Technology, Beijing 100081,
China}

\author{Xufeng Kou}

\affiliation{Department of Electrical Engineering, University of California, Los Angeles, California 90095, USA}

\author{Yabin Fan}

\affiliation{Department of Electrical Engineering, University of California, Los Angeles, California 90095, USA}

\author{Kang L. Wang}

\affiliation{Department of Electrical Engineering, University of California, Los Angeles, California 90095, USA}

\author{Yugui Yao}

\email{ygyao@bit.edu.cn}

\affiliation{School of Physics, Beijing Institute of Technology, Beijing 100081,
China}

\affiliation{Beijing National Laboratory for Condensed Matter Physics and Institute
of Physics,\linebreak{}
 Chinese Academy of Sciences, Beijing 100190, China}


\begin{abstract}
Magnetic interaction with the gapless surface states in topological insulator (TI) has
been predicted to give rise to a few exotic quantum phenomena.
However, the effective magnetic doping of TI is still challenging in experiment.
Using first-principles calculations, the magnetic doping properties (V, Cr, Mn
and Fe) in three strong TIs (Bi$_{2}$Se$_{3}$, Bi$_{2}$Te$_{3}$ and Sb$_{2}$Te$_{3}$) are investigated.
We find that for all three TIs the cation-site
substitutional doping is most energetically favorable with anion-rich
environment as the optimal growth condition. Further our results show
that under the nominal doping concentration of 4\%, Cr and Fe doped Bi$_{2}$Se$_{3}$,
Bi$_{2}$Te$_{3}$, and Cr doped Sb$_{2}$Te$_{3}$ remain as insulator, while all TIs doped
with V, Mn and Fe doped Sb$_{2}$Te$_{3}$ become metal. We also show that the magnetic
interaction of Cr doped Bi$_{2}$Se$_{3}$ tends to be ferromagnetic, while Fe doped
Bi$_{2}$Se$_{3}$ is likely to be antiferromagnetic.
Finally, we estimate the magnetic coupling and the Curie temperature
for the promising ferromagnetic insulator (Cr doped Bi$_{2}$Se$_{3}$)
by Monte Carlo simulation. These findings may provide important
guidance for the magnetism incorporation in TIs experimentally.
\end{abstract}


\maketitle

\section{INTRODUCTION}

In recent years, topological insulators characterized by insulating
bulk states and gapless conducting surface states have been studied intensively both theoretically and experimentally. \cite{Bernevig2006,Moore2007,Fu2007a,Zhang2009,Xia2009,Hsieh2009,Hasan2010,Qi2011_RevModPhys}
Specifically tetradymite compounds Bi$_{2}$Se$_{3}$, Bi$_{2}$Te$_{3}$
and Sb$_{2}$Te$_{3}$ are found to be three-dimensional strong topological insulators
with realistically large (a few hundred meV) bulk gap and simple surface
electronic structure. \cite{Zhang2009,Xia2009,Chen2009} On the other hand, even before
the concept of topological insulator, great efforts were made
to incorporate magnetism into these systems for potential
diluted magnetic semiconductors (DMS). For example, ferromagnetism
was reported in Cr doped Bi$_{2}$Se$_{3}$,\cite{Haazen2012} Mn
and Fe doped Bi$_{2}$Te$_{3}$ \cite{Choi2005,Kulbachinskii2002,Choi2004,Hor2010d,Niu2011}
and V, Cr and Mn doped Sb$_{2}$Te$_{3}$. \cite{Dyck2002,Dyck2005,Chien2007,Choi2005,Choi2004}
The ferromagnetism in topological insulator will
break the time-reversal symmetry, this intricate interplay between topological order
and ferromagnetism aroused a few proposals to realize exotic quantum
phenomena, \cite{Qi2008,Yu2010b,Liu2009,Qi2009,Garate2010,Wray2011b,Jin2011,Zhang2013,Kim2013,Lang2013} such
as, magnetoelectric effect \cite{Qi2008} and quantum anomalous Hall effect (QAHE). \cite{Yu2010b}
Experimentally the massive Dirac fermion spectrum was reported in both Mn and Fe doped Bi$_{2}$Se$_{3}$
surface, \cite{Chen2010i} complex spin texture was revealed in Mn doped Bi$_{2}$Te$_{3}$ and QAHE was recently observed in Cr$_{0.15}$(Bi$_{0.1}$Sb$_{0.9}$)$_{1.85}$Te$_{3}$
film under 30 mK.~\cite{Chang2013}

However, in experiment it is still challenging to incorporate
stable ferromagnetism in the TIs aforementioned. For example, ferromagnetism
in Fe doped Bi$_{2}$Te$_{3}$ and Sb$_{2}$Te$_{3}$ is hardly detected
even in low temperature. \cite{Zhou2006,Chien2007,He2011} For Bi$_{2}$Se$_{3}$
with Mn doping a spin glass state rather than ferromagnetic state
is observed.~\cite{Choi2005} Also both antiferromagnetism~\cite{Choi2011}
and ferromagnetism~\cite{Haazen2012,Kou2012} were observed in Cr doped Bi$_{2}$Se$_{3}$.
The similar controversy also exists from different groups for Fe doped
Bi$_{2}$Se$_{3}$. \cite{Kulbachinskii2002,Sugama2001,Salman2012,Choi2011}
This may be related to different magnetic atoms distribution within
the host material caused by the sample preparation, such as, temperature,
flux ratio, and chemical potentials of constituent atoms.

In order to clarify this issue, we systematically investigate the stability,
electronic and magnetic properties of 3$d$ transition metal (TM)
elements V, Cr, Mn and Fe doped Bi$_{2}$Se$_{3}$, Bi$_{2}$Te$_{3}$
and Sb$_{2}$Te$_{3}$ using DFT calculations and Monte Carlo simulations.
We first assess the feasibility of magnetic doping in Bi$_{2}$Se$_{3}$,
Bi$_{2}$Te$_{3}$ and Sb$_{2}$Te$_{3}$ under different growth environment
according to formation energy calculations. \cite{Zhang1991,VandeWalle2004}
The preferred site for the doping magnetic atoms and the optimal growth
conditions are identified. Further the electronic band structure results
show that Cr and Fe doped Bi$_{2}$Se$_{3}$, Bi$_{2}$Te$_{3}$,
and Cr doped Sb$_{2}$Te$_{3}$ remain as magnetic insulator with
substantially reduced band gap , while all TIs doped with V and Mn
as well as Fe doped Sb$_{2}$Te$_{3}$ become magnetic metal.
Additionally the magnetic coupling strength between
magnetic atoms is studied and Curie temperature for typical concentration is
estimated using Monte Carlo simulations.

This paper is organized as follows: In Sec. II we describe the method
for all the calculations proceeding. In Sec. III, we first identify
the native defects of Bi$_{2}$Se$_{3}$, Bi$_{2}$Te$_{3}$ and Sb$_{2}$Te$_{3}$,
which may be responsible for the intrinsic non-insulating bulk states
observed in experiment. Then we calculate the formation energies, electronic and
magnetic properties for magnetic atom doped TIs. We
additionally show the Monte Carlo simulations for the estimation of
magnetic coupling strength and Curie temperature. Finally we conclude
our paper with a brief summary of those findings.

\section{Method}

\begin{figure}
\begin{centering}
\includegraphics[width=3.4in]{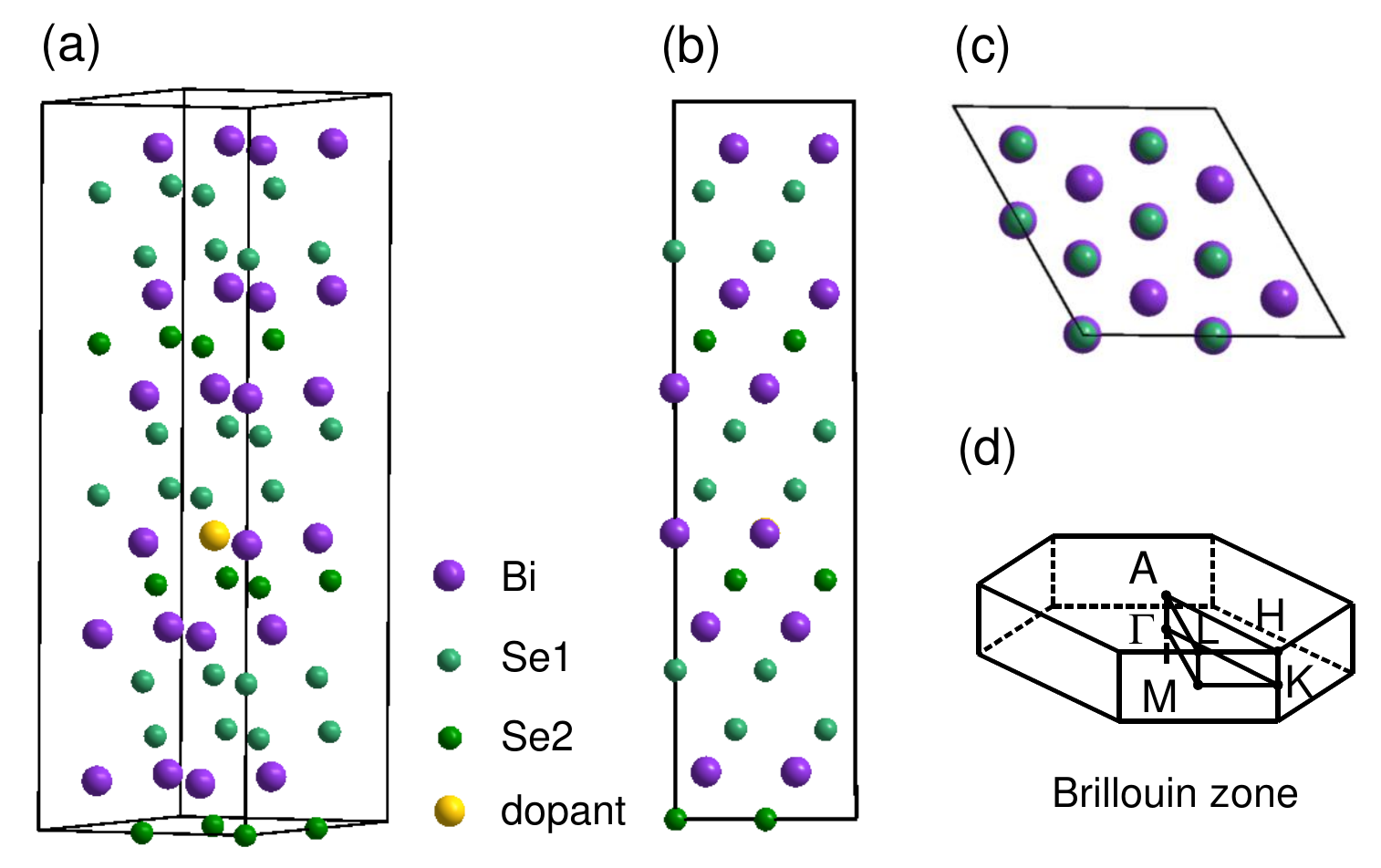}
\par\end{centering}

\centering{}\caption{(color online). (a) Crystal structure Illustration of a $2\times2\times1$
supercell for modeling a single dopant in bulk Bi$_{2}$Se$_{3}$,
with (b) for side view and (c) for top view. (d) Brillouin zone and
high symmetry points of the $2\times2\times1$ supercell. There are
two types of Se atoms, including Se1 involving the van der Waals bonding
and the other Se2.}
\label{fig:crystal}
\end{figure}

All the first-principles calculations are performed using projected augmented wave
(PAW) \cite{Blochl1994} potentials with Perdew-Burke-Ernzerhof type
generalized gradient approximation (GGA) \cite{Perdew1996} as implemented
in the Vienna \textit{ab initio} simulation package (VASP). \cite{Kresse1993a,Kresse1996}
In particular, spin orbit coupling (SOC) is explicitly included due to
the strong relativistic effect in Bi and Sb elements, and the significant
impact on electronic structure and formation energy, \cite{SOCnote} as
also revealed by West \textit{et al}.~\cite{West2012a} We choose hexagonal cell
with the experimental lattice constants a=4.138 {\AA}, c=28.64 {\AA} for Bi$_{2}$Se$_{3}$;
a=4.383 {\AA}, c=30.487 {\AA} for Bi$_{2}$Te$_{3}$ and a=4.250 {\AA}, c=30.35 {\AA} for Sb$_{2}$Te$_{3}$.
The cutoff energy for the plane wave expansion
of electron wavefunction was set at 300 eV. A gamma-centered $7\times7\times2$
k-mesh was adopted to sample the Brillouin zone for $2\times2\times1$
Bi$_{2}$Se$_{3}$, Bi$_{2}$Te$_{3}$ and Sb$_{2}$Te$_{3}$ supercells
as illustrated in Fig.~\ref{fig:crystal}. As calculating the energies of charged defects/dopants,
a jellium background charge is added. All atoms in each doped supercell
are fully relaxed through the conjugate-gradient algorithm until
the residual force on each atom is less than 0.02 eV/{\AA}.
The numerical errors of calculated formation energy are controlled to be less than 20 meV.

The formation energy $\Delta H_{f}(D,\, q)$ of defect or impurity
$D$ in charge state $q$ as a function of the Fermi energy $E_{F}$
and the chemical potential $\mu_{i}^{a}$ of atom $i$ is defined as~\cite{VandeWalle2004}

\begin{align}
\Delta H_{f}(D,\, q) & =E_{tot}(D,\, q)-E_{tot}(bulk)-\sum_{i}n_{i}\mu_{i}^{a}\nonumber \\
 & +q(E_{F}+E_{V}+\triangle V),\label{eq:formation energy formula}
\end{align}
 where $E_{tot}(D,\, q)$ (defect+host) is the total energy of a supercell
of host material with one defect or impurity $D$ in charge state
$q$, and $E_{tot}(bulk)$ (host only) is the total energy of the
equivalent supercell containing only pure host. $\mu_{i}^{a}$ denotes
the chemical potential for species $i$ (host atoms or dopants), and
$n_{i}$ indicates the corresponding number that have been added to
($n_{i}>0$) or removed from ($n_{i}<0$) the supercell. Here, it
is noted that $\mu_{i}^{a}$ is given with respect to the value of solid
phase $\mu_{i}^{solid}$, i.e., the absolute value of the chemical
potential $\mu_{i}^{a}=\mu_{i}+\mu_{i}^{solid}$. $E_{F}$
is the Fermi energy, referenced to the valence band maximum (VBM)
of the pure host crystal $E_{V}$. $\triangle V$ is a potential alignment
due to different energy references in defect containing supercell
and pure supercell in DFT calculations.

The chemical potentials depend on the experimental growth condition.
The values of $\mu_{i}$
are determined as follows, as taking Bi$_{2}$Se$_{3}$ for example,
first, $\mu_{\textrm{Bi}}\leq0$ and $\mu_{\textrm{Se}}\leq0$ to avoid precipitation
of solid elements. To maintain equilibrium growth of Bi$_{2}$Se$_{3}$,
it requires $2\mu_{\textrm{Bi}}+3\mu_{\textrm{Se}}=\Delta H_{f}(\textrm{Bi}_{2}\textrm{Se}_{3})$. Here,
$\Delta H_{f}$(Bi$_{2}$Se$_{3}$) is the formation energy of Bi$_{2}$Se$_{3}$.
Furthermore, $x\mu_{M}+y\mu_{\textrm{Se}}\leqslant\Delta H_{f}(M{}_{x}\textrm{Se}_{y})$
to ensure that the competing phases $M{}_{x}\textrm{Se}_{y}$ can not precipitate,
where $M$ is the dopant atom, i.e., V, Cr, Mn and Fe in the paper.

$\Delta H_{f}(D,\, q)$ is a function of charge q and Fermi energy,
then we can determine the transition energy as the Fermi energy at
which $\Delta H_{f}(D,\, q)=\Delta H_{f}(D,\, q')$, i.e., where the
charge state of defect $D$ spontaneously transforms from $q$ to
$q'$. The concentration $c$ of defects or dopants at growth temperature
$T_{g}$ under thermodynamic equilibrium can be estimated from~\cite{VandeWalle2004}
\begin{equation}
c=Nexp(-\triangle H_{f}/k_{B}T_{g}),\label{eq:concentration}
\end{equation}
 where N is the number of sites that can be occupied in the lattice
(per unit volume), $\triangle H_{f}$ is defined in Eq.(\ref{eq:formation energy formula})
and $k_{B}$ is Boltzmann constant.

\section{Results and Discussions}

\subsection{Native defects}

\begin{figure*}
\begin{centering}
\includegraphics{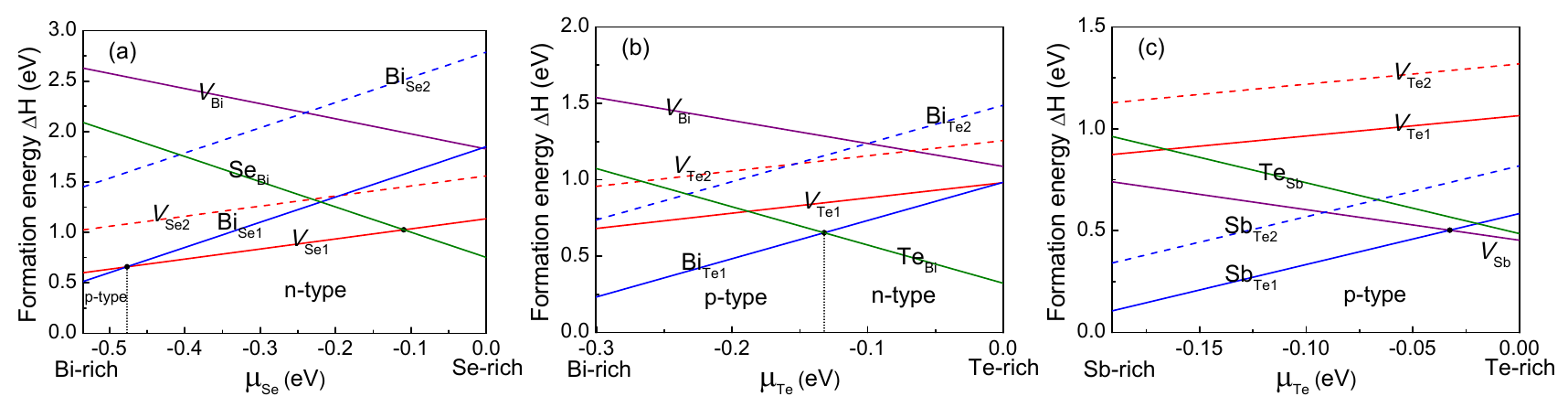}
\par\end{centering}

\centering{}\caption{(color online). The formation energy $\triangle H$ as a function
of anion chemical potential for all the possible intrinsic defects
in (a) Bi$_{2}$Se$_{3}$, (b) Bi$_{2}$Te$_{3}$ and (c) Sb$_{2}$Te$_{3}$.
$V_{\textrm{Bi}}$, $V_{\textrm{Sb}}$, $V_{\textrm{Se}}$ and $V_{\textrm{Te}}$ stand for bismuth vacancy,
antimony vacancy, selenium vacancy and tellurium vacancy, respectively,
while Bi$_{\textrm{Se}}$, Bi$_{\textrm{Te}}$, Sb$_{\textrm{Te}}$, Se$_{\textrm{Bi}}$ and Te$_{\textrm{Bi}}$
are antisites defects. $1$ and $2$ are labeled to distinguish Se
(Te) in different layers. $X$-rich
($X$ for Bi, Sb, Se or Te) indicate the extreme growth condition with
$\mu_{X}$ taking the maximum value in Eq.(\ref{eq:formation energy formula}).
 Vertically dotted lines highlight the boundary
of carrier types. }
\label{fig:formation-native}
\end{figure*}

\begin{table*}[!htbp]
\caption{A list of theoretical and experimental reports of carrier type tendency in pure Bi$_{2}$Se$_{3}$, Bi$_{2}$Te$_{3}$ and Sb$_{2}$Te$_{3}$.}

\medskip{}
\begin{ruledtabular}

\begin{tabular}{cccccccc}
\multirow{2}{*}{~}  & \multicolumn{3}{c}{$n$-type tendency} &  & \multicolumn{3}{c}{$p$-type tendency}\tabularnewline
\cline{2-4} \cline{6-8}
 & \multirow{1}{*}{Theory} &  & Experiment &  & Theory &  & Experiment\tabularnewline
\hline
Bi$_{2}$Se$_{3}$  & Our%
\footnote{most growth conditions,%
}, Ref. \onlinecite{West2012a,Wang2012-arXiv}  &  & Ref. \onlinecite{Xia2009,Hor2009,Hsieh2009,Wang2010,Urazhdin2004}  &  & Our%
\footnote{extremely Bi-rich condition,%
}, Ref. \onlinecite{Wang2012-arXiv} &  & \tabularnewline
Bi$_{2}$Te$_{3}$  & Our%
\footnote{Te-rich condition,%
}, Ref. \onlinecite{West2012a,Hashibon2011}  &  & Ref. \onlinecite{Chen2009,Giani1999,Yoo2005,Wang2011b,Urazhdin2004}  &  & Our%
\footnote{Bi-rich condition.%
}, Ref. \onlinecite{West2012a,Hashibon2011}  &  & Ref. \onlinecite{Chien2007,Lee2008,Wang2011b}\tabularnewline
Sb$_{2}$Te$_{3}$  &  &  &  &  & Our, Ref. \onlinecite{West2012a}  &  & Ref. \onlinecite{Giani1999,Gasenkova2001,Jiang2012}\tabularnewline
\end{tabular}

\end{ruledtabular}
\label{table:native-defect}
\end{table*}

Experimentally, Bi$_{2}$Se$_{3}$, Bi$_{2}$Te$_{3}$ and Sb$_{2}$Te$_{3}$
are always dominated by conducting bulk carriers rather than being
insulating even though they are all intrinsically narrow-band semiconductors.
This is related to the unintentional doping induced by native defects.
Bi$_{2}$Se$_{3}$ often shows $n$-type conductivity and is difficult to be
tuned into $p$-type via compensation doping,~\cite{Xia2009,Hor2009,Hsieh2009,Wang2010,Urazhdin2004}
while Sb$_{2}$Te$_{3}$ shows strong $p$-type tendency.~\cite{Giani1999,Gasenkova2001,Jiang2012}
For Bi$_{2}$Te$_{3}$, it is reported to be either $n$-type~\cite{Giani1999,Yoo2005,Chen2009,Urazhdin2004}
or $p$-type~\cite{Chien2007,Lee2008} depending on the growth method and environment.
We then identify how the carrier type varies with the chemical potentials.
The most possible native point defects including atom vacancies and antisites defects are considered.
The formation energy versus chemical potential is plotted in Fig.~\ref{fig:formation-native}.

As shown in Fig.~\ref{fig:formation-native}(a), donor-like defects $V_{\textrm{Se1}}$ and Se$_{\textrm{Bi}}$ dominate
in Bi$_{2}$Se$_{3}$ in the most range of growth conditions according to
their lowest formation energies among all the native point defects, as also revealed by Ref.~\onlinecite{Wang2012-arXiv,West2012a}.
This will lead to an intrinsic $n$-type doping as experimentally observed. In the extreme Bi-rich condition,
acceptor-like defect Bi$_{\textrm{Se1}}$ will be preferred and the resulting
doping will be $p$-type.

For Bi$_{2}$Te$_{3}$ in Fig.~\ref{fig:formation-native}(b), antisite defects Bi$_{\textrm{Te1}}$
and Te$_{\textrm{Bi}}$ are more preferred than other native point defects.
We identify that acceptor-like Bi$_{\textrm{Te1}}$ is likely to appear in
Bi-rich condition and donor-like Te$_{\textrm{Bi}}$ in Te-rich condition,
leading Bi$_{2}$Te$_{3}$ to be intrinsic $p$-type and intrinsic
$n$-type, respectively. Our result explains the experimentally reported
native $n$-$p$ amphoteric type conductivity of Bi$_{2}$Te$_{3}$.~\cite{Giani1999,Yoo2005,Chen2009,Urazhdin2004,Chien2007,Lee2008}
Our result agrees with Ref.~\onlinecite{West2012a}, while calculation without the inclusion of SOC
gives rather different values of formation energy.~\cite{Hashibon2011} Experimentally Ref.~\onlinecite{Wang2011b} reported
the co-existence of Te$_{\textrm{Bi}}$ antisite and Bi$_{\textrm{Te1}}$ antisite, rendering
Bi$_{2}$Te$_{3}$ to be either $n$-type tendency or $p$-type tendency. This result
further confirms our predication.

For Sb$_{2}$Te$_{3}$ in Fig.~\ref{fig:formation-native}(c), we find that antisite defect Sb$_{\textrm{Te1}}$ is dominant
with the lowest formation energy in most range of the growth conditions especially in Sb-rich condition.
This can be explained qualitatively by the similar atomic sizes of Sb atom and Te atom.
As the growth atmosphere evolves to be extremely Te-rich, antimony vacancy $V_{\textrm{Sb}}$ becomes
to be the most energetically stable. Note that both Sb$_{\textrm{Te1}}$ and $V_{\textrm{Sb}}$
are acceptor-like defects, Sb$_{2}$Te$_{3}$ is thus always intrinsic
$p$-type.~\cite{West2012a,Gasenkova2001,Jiang2012} Our results provide an important guidance
to carrier tuning in Bi$_{2}$Se$_{3}$, Bi$_{2}$Te$_{3}$ and Sb$_{2}$Te$_{3}$ as well as
intrinsic carrier environments for magnetic doping. Meanwhile, these findings also provide
a clear explanation to experimental reports, as listed in Table \ref{table:native-defect}.

\subsection{Formation energies of magnetic doping in Bi$_{2}$Se$_{3}$, Bi$_{2}$Te$_{3}$
and Sb$_{2}$Te$_{3}$}

\begin{figure*}
\begin{centering}
\includegraphics[scale=0.5]{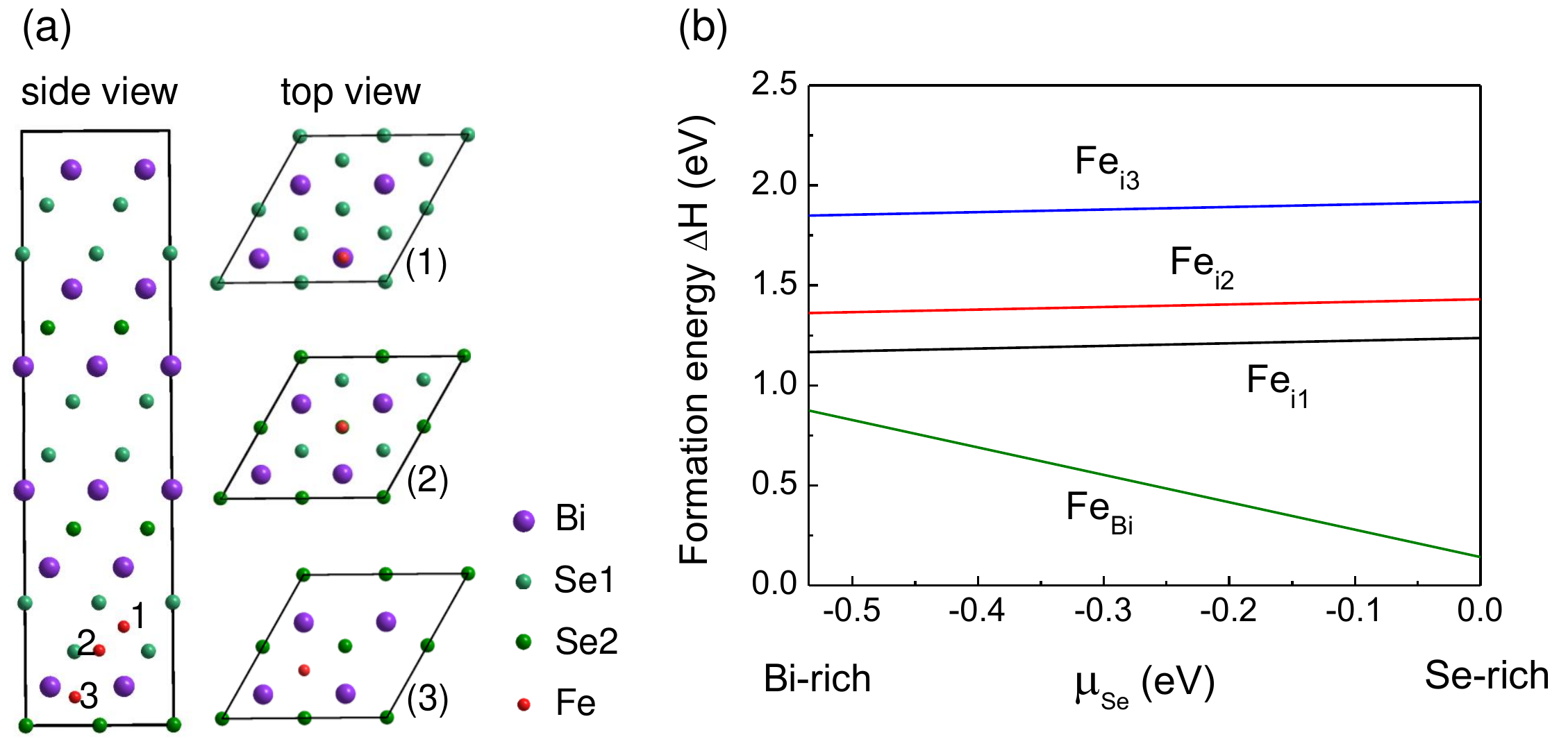}
\par\end{centering}

\centering{}\caption{(color online). (a) Possible interstitial sites for Fe in Bi$_{2}$Se$_{3}$
after relaxation. (b) The formation energy $\triangle H$ as a function
of anion chemical potential ($\mu_{\textrm{Se}}$) for interstitial Fe and
Fe substitutional for Bi in Bi$_{2}$Se$_{3}$. Fe$_{i1}$, Fe$_{i2}$,
and Fe$_{i3}$ stand for different sites for interstitial Fe in (a),
while Fe$_{\textrm{Bi}}$ denotes that Bi atom is doped by Fe atom.}
\label{fig:possible-interstitial}
\end{figure*}

\begin{figure*}
\begin{centering}
\includegraphics{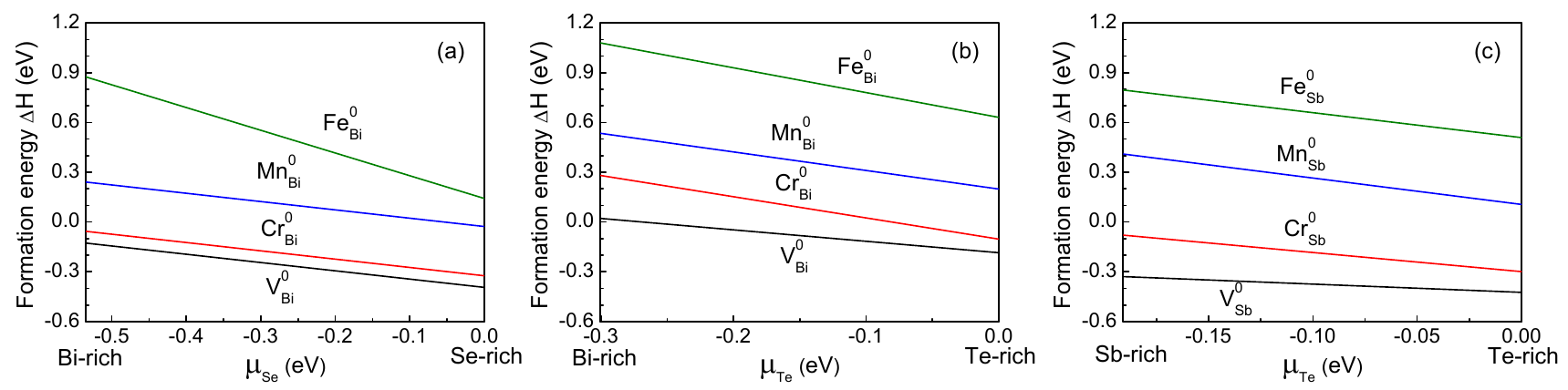}
\par\end{centering}

\centering{}\caption{(color online). Calculated formation energies of the most stable configurations
of single V, Cr, Mn, and Fe impurities doped (a) Bi$_{2}$Se$_{3}$,
(b) Bi$_{2}$Te$_{3}$ and (c) Sb$_{2}$Te$_{3}$ as a function of
the host element chemical potentials.}
\label{fig:formation-of-DM-chemical}
\end{figure*}

In this section, we will calculate the formation energies for the
incorporation of TM atoms into the three TIs. First, the site preference
of TM atom (substitutional or interstitial site) is studied. Notice that intercalated and interstitial
sites were reported to be preferred for Cu in Bi$_{2}$Se$_{3}$. \cite{Wang2011a}
We have considered all the possible interstitial sites in bulk Bi$_{2}$Se$_{3}$,
including interstitial sites between different layers (intercalated sites) and interstitial
sites on the same layer. Relaxed structures indicate that all the
interstitial atoms are relaxed to the three main sites, as shown in
Fig.~\ref{fig:possible-interstitial}(a). Formation energies for both substitution and interstitial cases are
shown in Fig.~\ref{fig:possible-interstitial}(b). We find that Bi substitutional site is
strongly preferred~\cite{West2012} regardless of the changes of growing
condition, as compared to all the possible interstitial sites. Our results
are well in line with recently experimental findings.~\cite{West2012,Song2012}
Then in the following we will mainly focus on cation substitutional doping.

Formation energy of TM doping as a function of chemical potential is shown in Fig.~\ref{fig:formation-of-DM-chemical}.
Similar to TM doping in Bi$_{2}$Se$_{3}$, \cite{Zhang2012}
the formation energies exhibit the same size effect in Bi$_{2}$Te$_{3}$
and Sb$_{2}$Te$_{3}$, that is, the formation energy is lowest
for V atom doping while highest for Fe atom doping. It is attributed to the
closest atom radius of V atom to the substituted Bi or Sb atom than other dopants.
Also we find that V and Cr have negative formation energies in Bi$_{2}$Se$_{3}$
and Sb$_{2}$Te$_{3}$ for the whole range of chemical potential,
indicating the doping of them can occur spontaneously. Recently, heavy Cr doping
of Bi$_{2}$Se$_{3}$ with the concentration up to 23$\%$ was reported~\cite{Liu2012}
and AFM measurement indicated Cr atoms of 20$\%$ doping concentration were uniformly
distributed.~\cite{Kou2012} For Sb$_{2}$Te$_{3}$,
even in Sb$_{1.41}$Cr$_{0.59}$Te$_{3}$, Cr atoms can homogeneously distribute
without clustering. \cite{Chien2007} However, in Bi$_{2}$Te$_{3}$,
only V can be spontaneously doped. Also it's rather different
that the formation energies of Mn and Fe doping are positive values
in all three TIs, suggesting the doping of them is not spontaneous
except Mn in Bi$_{2}$Se$_{3}$ at extremely Se-rich atmosphere. Indeed,
Hor \textit{et al}. showed that 9\% Mn can substitute for Bi atoms
with randomly distributing in Bi$_{2}$Te$_{3}$.~\cite{Hor2010d} Fe is confirmed
even more difficult to be doped in Bi$_{2}$Se$_{3}$ with the effective doping
concentration less than 2\%.~\cite{Cha2010} Notice that
Bi$_{2}$Te$_{3}$ has entire higher formation energies for all these dopants
than that in Bi$_{2}$Se$_{3}$ and Sb$_{2}$Te$_{3}$, suggesting it is relatively
more difficult to dope those atoms in Bi$_{2}$Te$_{3}$.

\begin{figure*}
\begin{centering}
\includegraphics{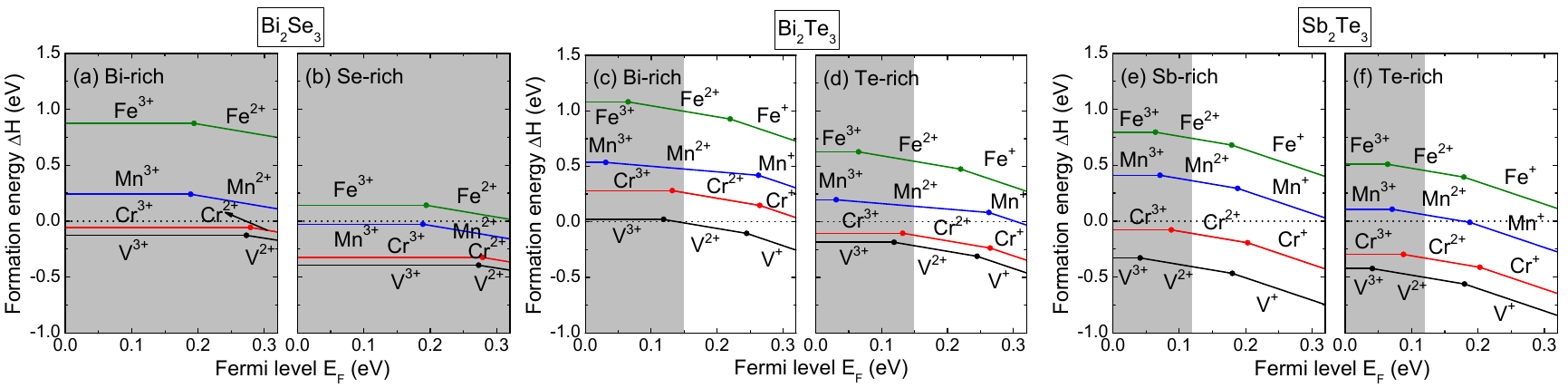}
\par\end{centering}

\centering{}\caption{(color online). Calculated defect formation energy $\triangle H$
as a function of Fermi energy $E_{F}$ for magnetic doped Bi$_{2}$Se$_{3}$
under (a) Bi-rich condition and (b) Se-rich condition; Bi$_{2}$Te$_{3}$
under (c) Bi-rich condition and (d) Te-rich condition and Sb$_{2}$Te$_{3}$
under (e) Sb-rich condition and (f) Te-rich condition. Here, $E_{F}$ is referenced
to the valence-band maximum ($E_{V}$) in the bulk. The shaded areas
highlight band gaps, where $E_{F}$ can range, as we mainly focus,
from VBM ($E_{F}=0$) to CBM ($E_{F}=Eg$). Calculated values of
band gaps with SOC for Bi$_{2}$Se$_{3}$, Bi$_{2}$Te$_{3}$ and
Sb$_{2}$Te$_{3}$ are 0.32, 0.15 and 0.12 eV, respectively. Solid
dots denote the defect transition energies between different charge
states.}
\label{fig:formation-of-DM-fermi}
\end{figure*}

From section III A, we have known that the native defects are responsible for the
various intrinsic carriers doping. Such carrier environment is expected to affect
the formation energy of magnetic dopant with nonzero charge state according
to Eq.(\ref{eq:formation energy formula}). Then we study the possible charge
states by calculating the formation energy as a function of Fermi energy,
as shown in Fig.~\ref{fig:formation-of-DM-fermi}.

(i) For Bi$_{2}$Se$_{3}$, Bi$_{2}$Te$_{3}$ and Sb$_{2}$Te$_{3}$,
anion-rich growth conditions (Se-rich or Te-rich) with lower formation
energies are revealed better than cation-rich conditions (Bi-rich
or Sb-rich) for magnetic atoms doping, which is consistent with experimental
reports.~\cite{Chien2007,Song2010}

(ii) For Bi$_{2}$Se$_{3}$, we find that as the Fermi energy $E_{F}$
ranges from VBM at 0.0 (left edge of the shaded area) to CBM at $E_{g}$
(right edge of the shaded area), V, Cr, Mn and Fe atoms are almost
stable with charge state of $+3$, i.e., neutral substitute for Bi
atoms, which indicate that dopants do not introduce free carriers
to the host materials. This result agrees with theoretical study
from Larson et al. \cite{a-Larson2008} and
has been experimentally confirmed in Cr doped Bi$_{2}$Se$_{3}$.~\cite{Kou2012}
Although, as $E_{F}$ shifts very close to CBM, i.e.,
under extremely $n$-type condition, dopants tend to act as acceptors
with valency$+2$, especially Mn and Fe. Experimentally, Mn was indeed found
to show hole doping effect in Bi$_{2}$Se$_{3}$.~\cite{Chen2010i}

(iii) For Bi$_{2}$Te$_{3}$, the formation energies of TMs are larger
than that in Bi$_{2}$Se$_{3}$ or Sb$_{2}$Te$_{3}$ both at Bi-rich and
Te-rich conditions. Mn and Fe can be neutral doped in very $p$-type
conditions. Mostly, Mn tends to act as an acceptor (Mn$_{\textrm{Bi}}^{-}$)
with valence state Mn$^{2+}$in Bi$_{2}$Te$_{3}$. This agrees well
with the experimental result from Hor \textit{et al}.~\cite{Hor2010d}

(iv) For Sb$_{2}$Te$_{3}$, a similar size effect among V, Cr, Mn
and Fe dopants is observed. V and Cr with negative formation energies
can be spontaneously incorporation under Te-rich atmosphere. Cr can
be neutrally doped (Cr$_{\textrm{Sb}}^{0}$)~\cite{Dyck2005,Chien2007} in the
most range of Fermi level.

\begin{figure}
\begin{centering}
\includegraphics[scale=0.15]{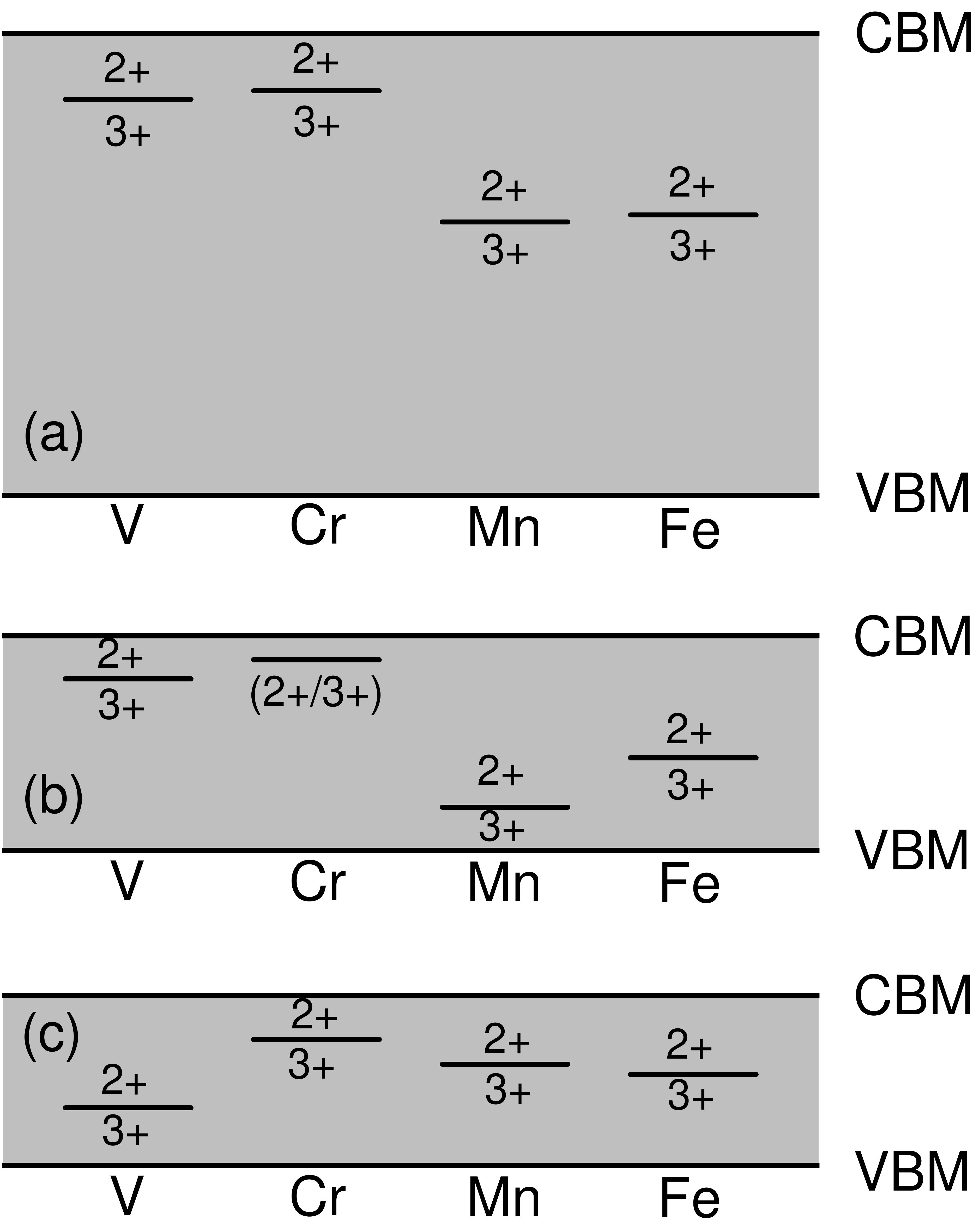}
\par\end{centering}

\centering{}\caption{Schematic representation of the thermodynamic transition levels for
magnetic doped (a) Bi$_{2}$Se$_{3}$, (b) Bi$_{2}$Te$_{3}$, and
(c) Sb$_{2}$Te$_{3}$, which corresponding to the solid dots in Fig.~\ref{fig:formation-of-DM-fermi}.
The transition levels shown are referenced to valence-band maximum
(VBM) in the bulk. The distances of $\varepsilon(3+/2+)$
to VBM, for example, indicate the thermal ionization energy of simple acceptors,
respectively. The shaded area highlight band gaps of Bi$_{2}$Se$_{3}$, Bi$_{2}$Te$_{3}$
and Sb$_{2}$Te$_{3}$, as shown in Fig.~\ref{fig:formation-of-DM-fermi}.}
\label{fig:formation-of-DM-transition}
\end{figure}

%

\begin{figure*}
\begin{centering}
\includegraphics[scale=0.25]{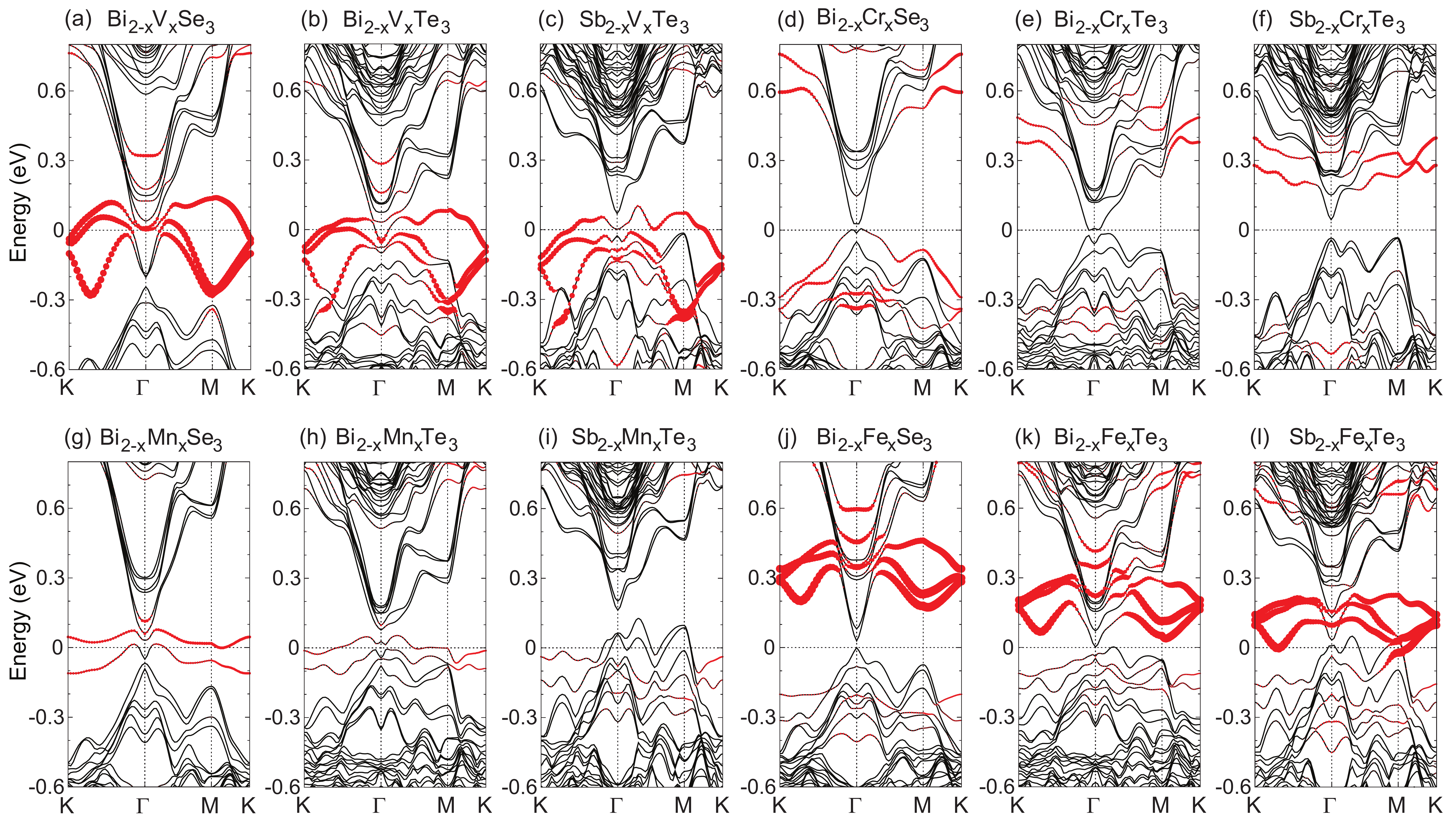}
\par\end{centering}

\caption{
(color online). SOC band structures for relaxed V, Cr, Mn and Fe doped Bi$_{2}$Se$_{3}$,
Bi$_{2}$Te$_{3}$ and Sb$_{2}$Te$_{3}$ with the nominal doping concentration of 4$\%$ (x=0.083).
The size of red dots denotes the contribution of TM-\textit{d} states.
         }
\label{fig:soc-band}
\end{figure*}

\begin{table*}
\caption{Relaxed band gaps and magnetic moments ($m$) of Cr and Fe doped Bi$_{2}$Se$_{3}$,
Bi$_{2}$Te$_{3}$ and Sb$_{2}$Te$_{3}$ with SOC. The results of
GGA+U with U=3 eV and J=0.87 eV are also listed. }

\medskip{}
 \begin{ruledtabular}

\begin{tabular}{ccccccccccc}
\multirow{2}{*}{System}  &  & \multicolumn{4}{c}{Cr} &  & \multicolumn{4}{c}{Fe}\tabularnewline
\cline{3-6} \cline{8-11}
 &  & gap  & gap (+U)  & $m$  & $m$ (+U)  &  & gap  & gap (+U)  & $m$  & $m$ (+U)\tabularnewline
\hline
Bi$_{2}$Se$_{3}$  &  & 0.010 eV  & 0.025 eV  & 2.94 $\mu_{B}$  & 2.99 $\mu_{B}$  &  & 0.028 eV  & 0.028 eV  & 4.73 $\mu_{B}$  & 4.99 $\mu_{B}$\tabularnewline
Bi$_{2}$Te$_{3}$  &  & 0.017 eV  & 0.019 eV  & 2.93 $\mu_{B}$  & 2.98 $\mu_{B}$  &  & 0.0024 eV  & 0.041 eV  & 4.17 $\mu_{B}$  & 4.09 $\mu_{B}$\tabularnewline
Sb$_{2}$Te$_{3}$  &  & 0.077 eV  & 0.100 eV  & 3.06 $\mu_{B}$  & 3.16 $\mu_{B}$  &  & 0  & 0  & 4.17 $\mu_{B}$  & 4.03 $\mu_{B}$\tabularnewline
\end{tabular}

\end{ruledtabular}
\label{table:gap-moment}
\end{table*}

From Fig.~\ref{fig:formation-of-DM-fermi} we can easily determine the thermodynamic transition level
from one charge state to another for different dopants, which can
be observed in deep-level transient spectroscopy (DLTS) experiments
or temperature-dependent Hall measurements. \cite{VandeWalle2004}
We schematically illustrate them in Fig.~\ref{fig:formation-of-DM-transition}. Bi$_{2}$Se$_{3}$ is
revealed above to be mostly intrinsic $n$-type, nevertheless, from Fig.~\ref{fig:formation-of-DM-transition}(a)
we can see that V, Cr, Mn and Fe are neutral stable in most carrier
environment for Bi$_{2}$Se$_{3}$, even in $n$-type condition ($E_{F}$
near to CBM). From Fig.~\ref{fig:formation-of-DM-transition}(b) for Bi$_{2}$Te$_{3}$, it is indicated that
under $n$-type condition, only V and Cr can neutrally substitute for Bi
atom, while Mn and Fe are energetically stable with charge state of
$+2$. Conversely, neutral substituting is more likely to appear by
V, Cr and Fe doping than Mn in $p$-type Bi$_{2}$Te$_{3}$. From Fig.~\ref{fig:formation-of-DM-transition}(c)
for Sb$_{2}$Te$_{3}$, as compared to V, Mn and Fe, Cr$_{\textrm{Sb}}$ is
especially deep and more difficult to be ionized from charge state
Cr$^{3+}$ to Cr$^{2+}$, suggesting Cr is the best candidate atom
for the realization of QAHE in Sb$_{2}$Te$_{3}$. Actually, Cr has
already been experimentally confirmed can substitute for Sb with Cr$^{3+}$in
Sb$_{2}$Te$_{3}$. \cite{Dyck2005,Chien2007}

\subsection{Electronic structure of magnetically doped Bi$_{2}$Se$_{3}$, Bi$_{2}$Te$_{3}$
and Sb$_{2}$Te$_{3}$}

To elaborate the electronic properties with magnetic atoms introduced
in Bi$_{2}$Se$_{3}$, Bi$_{2}$Te$_{3}$ and Sb$_{2}$Te$_{3}$,
we further calculate the band structures for all those magnetically
doped TIs and they are shown in Fig.~\ref{fig:soc-band}.
In Fig.~\ref{fig:soc-band}, additional states appear in the band gaps of TM doped Bi$_{2}$Se$_{3}$, Bi$_{2}$Te$_{3}$ and Sb$_{2}$Te$_{3}$ and give rise to semiconducting or metallic ground states, comparing to the pure host materials.~\cite{Zhang2009,Zhang2012,Zhang2010} The plots of TM-\textit{d} orbital projected band structures show the states near the band gaps are from sizable hybridization between TM states and p states of host materials, most obvious at the Gamma point of the Brillouin zone. The results indicate that V and Mn doped Bi$_{2}$Se$_{3}$, Bi$_{2}$Te$_{3}$ and Sb$_{2}$Te$_{3}$ are metals, as shown in Figs.~\ref{fig:soc-band}(a)-\ref{fig:soc-band}(c) and Figs.~\ref{fig:soc-band}(g)-\ref{fig:soc-band}(i), respectively.
However, Cr doped Bi$_{2}$Se$_{3}$ exhibits an insulating magnetic state
with the energy gap 0.01 eV. Compared to pure Bi$_{2}$Se$_{3}$,~\cite{Zhang2012}
inverted bands are remain observed in the doped system, indicating the
topological nontrivial property. Accounting to our calculations, we get
similar results in Cr doped Bi$_{2}$Te$_{3}$ with the band gap of
0.017 eV and Cr doped Sb$_{2}$Te$_{3}$ with a larger band gap of
0.077 eV. From Figs.~\ref{fig:soc-band}(j)-(l), we find that
Fe doped Bi$_{2}$Se$_{3}$ manifests insulating behavior
with the band gap of 0.028 eV, whereas magnetic moments of Fe doped
Bi$_{2}$Te$_{3}$ and Sb$_{2}$Te$_{3}$ are less than 5 $\mu_{B}$,
rendering Fe doped Bi$_{2}$Te$_{3}$ to be semi-metal with a narrow
gap 0.0024 eV and Fe doped Sb$_{2}$Te$_{3}$ to be gapless. The phenomenon of gap closing may lead to a topological phase transition.~\cite{Jin2011,Zhang2013,Kim2013}
The resulting values of band gaps and magnetic moments upon doping are listed in
Table \ref{table:gap-moment}. In order to investigate the effect of electron-electron
correlation on band gap and magnetic moment, we further perform GGA+U
calculations with U ranging from 3 to 6 eV and J=0.87 eV. We find only
slight modification of the band gaps and magnetic moments.

\begin{figure}
\begin{centering}
\includegraphics[scale=0.5]{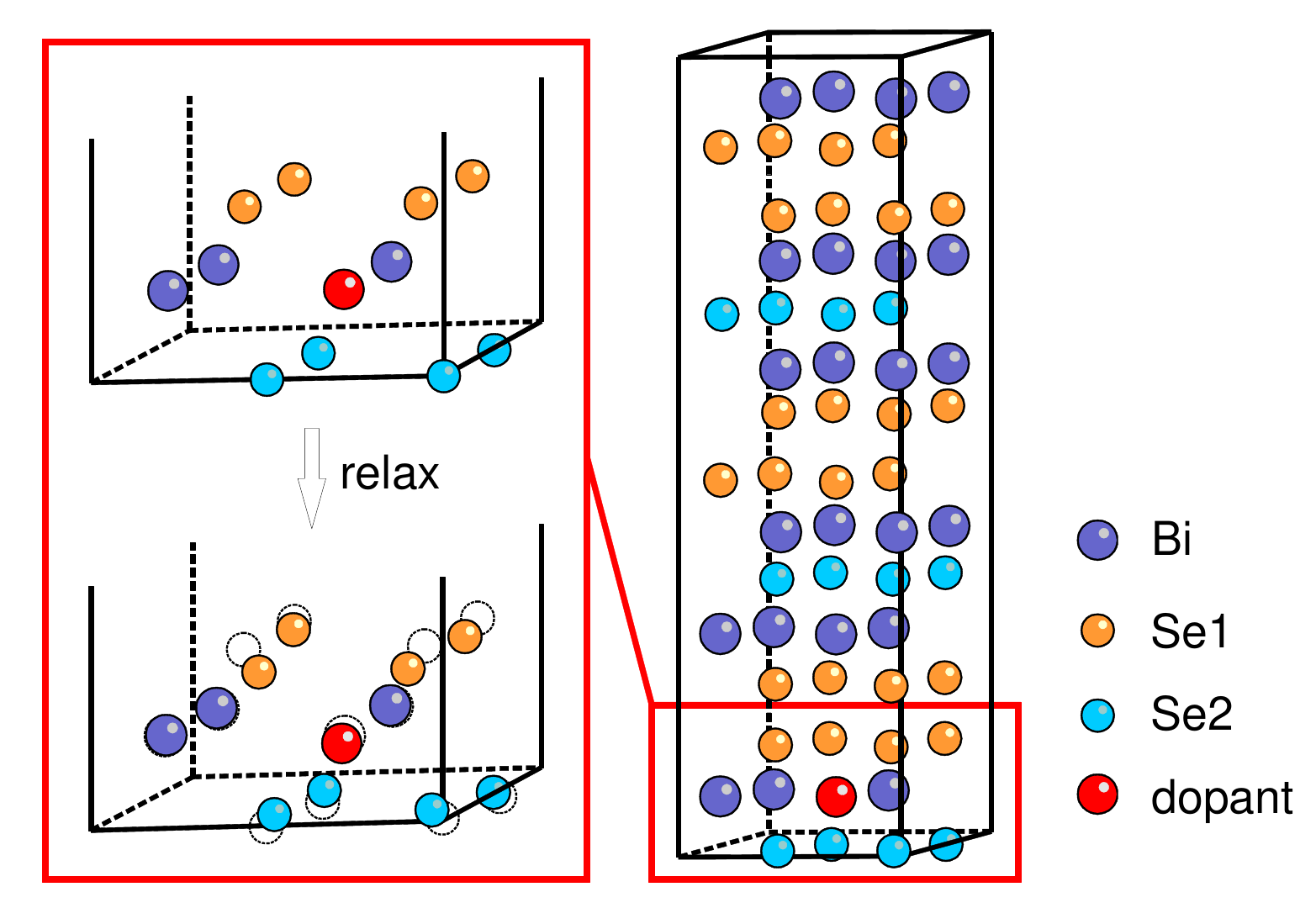}
\linebreak{}

\par\end{centering}

\centering{}\caption{(color online). Dynamics illustration of the magnetically doped Bi$_{2}$Se$_{3}$.
Note that those Se atoms neighboring to the dopants tend to close
to the dopants, while Bi atoms hardly move.}
\label{fig:relax}
\end{figure}

Notice that the band gaps are 0.32 eV, 0.15 eV and 0.12 eV for Bi$_{2}$Se$_{3}$,
Bi$_{2}$Te$_{3}$ and Sb$_{2}$Te$_{3}$ respectively. However we find the band gaps are substantially
reduced to several meV upon doping. This result hints that
QAHE should be observed under low temperature in magnetically doped Bi$_{2}$Se$_{3}$
family, which is consistent with recent experimental reports.~\cite{Chang2013}
In order to uncover the reason which causes this band gap reduction,
we study the effect of structural relaxation on the band gap.
In Fig.~\ref{fig:relax}, we show the schematic structures of doped Bi$_{2}$Se$_{3}$ before
and after relaxation. The structural relaxation shows Se atoms neighboring
to dopants move inward to the dopants by sizable distances (See Table \ref{table:relax}), as
consistent with Ref. \onlinecite{a-Larson2008}.
This suggests that hybridization
between TM dopant and the neighboring Se will be strengthened and thus the
impurity bands may be broadened.~\cite{Zhang2012} Specifically
as reported in our previous paper, \cite{Zhang2012} calculated
band gaps of Cr and Fe doped Bi$_{2}$Se$_{3}$ without relaxation are 0.28 and 0.18 eV, respectively.
While after structural relaxation, the band gaps are reduced to 0.01 and 0.028 eV, respectively.
GGA+U calculation for the relaxation gives essentially the same
results shown in Table \ref{table:relax} as only GGA calculation. Additional SOC relaxations for Cr and Fe doped
Bi$_{2}$Se$_{3}$ suggest the relaxed distances change within only
the order of 0.001 {\AA} comparing to non-SOC cases. We thus conclude
that the band gap reduction is induced by Se-dopant hybridization.

\begin{table}[ptbh]
 \caption{Amplitude of the bond length variation $\triangle d$ (in {\AA{}})
from GGA and GGA+U calculations for Cr and Fe doped Bi$_{2}$Se$_{3}$ with all the atoms fully relaxed.
Here, $\triangle d=d_{relaxed}-d_{initial}$, where negative value
means a decrease of the bond length. }

\medskip{}
 \begin{ruledtabular}

\begin{tabular}{ccccccc}
\multirow{2}{*}{Bond length}  &  & \multicolumn{2}{c}{Cr} &  & \multicolumn{2}{c}{Fe}\tabularnewline
\cline{3-4} \cline{6-7}
 &  & $\triangle d$  & $\triangle d$ (+U)  &  & $\triangle d$  & $\triangle d$ (+U)\tabularnewline
\hline
\ $d_{dopant-Bi}$  &  & 0.009  & 0.009  &  & 0.000  & 0.002\tabularnewline
\ $d_{dopant-Se1}$  &  & -0.342  & -0.305  &  & -0.404  & -0.360\tabularnewline
\ $d_{dopant-Se2}$  &  & -0.390  & -0.362  &  & -0.239  & -0.263\tabularnewline
\end{tabular}

\end{ruledtabular}
\label{table:relax}
\end{table}

\subsection{Magnetic properties of magnetically doped Bi$_{2}$Se$_{3}$}

\begin{table}
\begin{centering}
\caption{Calculated magnetic coupling strength [$(E_{AFM}-E_{FM})/2$] between
two substitutional Cr atoms with different distances $d$ for Bi$_{2-x}$Cr$_{x}$Se$_{3}$ ($x=0.074$).
Two Bi atoms are replaced by dopant atoms at site $i$=0 and site $j$ ($j$= 1, 2, 3...) in a $3\times3\times1$ supercell of Bi$_{2}$Se$_{3}$, which gives distinct configurations. The first seven nearest neighbor configurations of Cr atoms are considered as listed. Farther neighbor configurations are ignored for their large distances of Cr atoms [$d$ of configuration (0, 8) is already larger than 9 {\AA}] and little contributions to the magnetic coupling strengths (less than 2 meV).}
\par\end{centering}

\medskip{}

\begin{ruledtabular}

\begin{centering}
\begin{tabular}{ccc}
Configuration ($i$, $j$) & $d$ ({\AA}) & $(E_{AFM}-E_{FM})/2$ (meV)\tabularnewline
\hline
(0, 1) & 4.138 & 22.0\tabularnewline
(0, 2) & 4.456 & 13.8\tabularnewline
(0, 3) & 5.785 & 8.7\tabularnewline
(0, 4) & 6.081 & -3.8\tabularnewline
(0, 5) & 7.113 & 5.5\tabularnewline
(0, 6) & 7.167 & 2.1\tabularnewline
(0, 7) & 7.355 & -1.8\tabularnewline
\end{tabular}
\par\end{centering}

\end{ruledtabular}
\label{table:magnetic-coupling}
\end{table}

\begin{figure}
\begin{centering}
\includegraphics[scale=0.35]{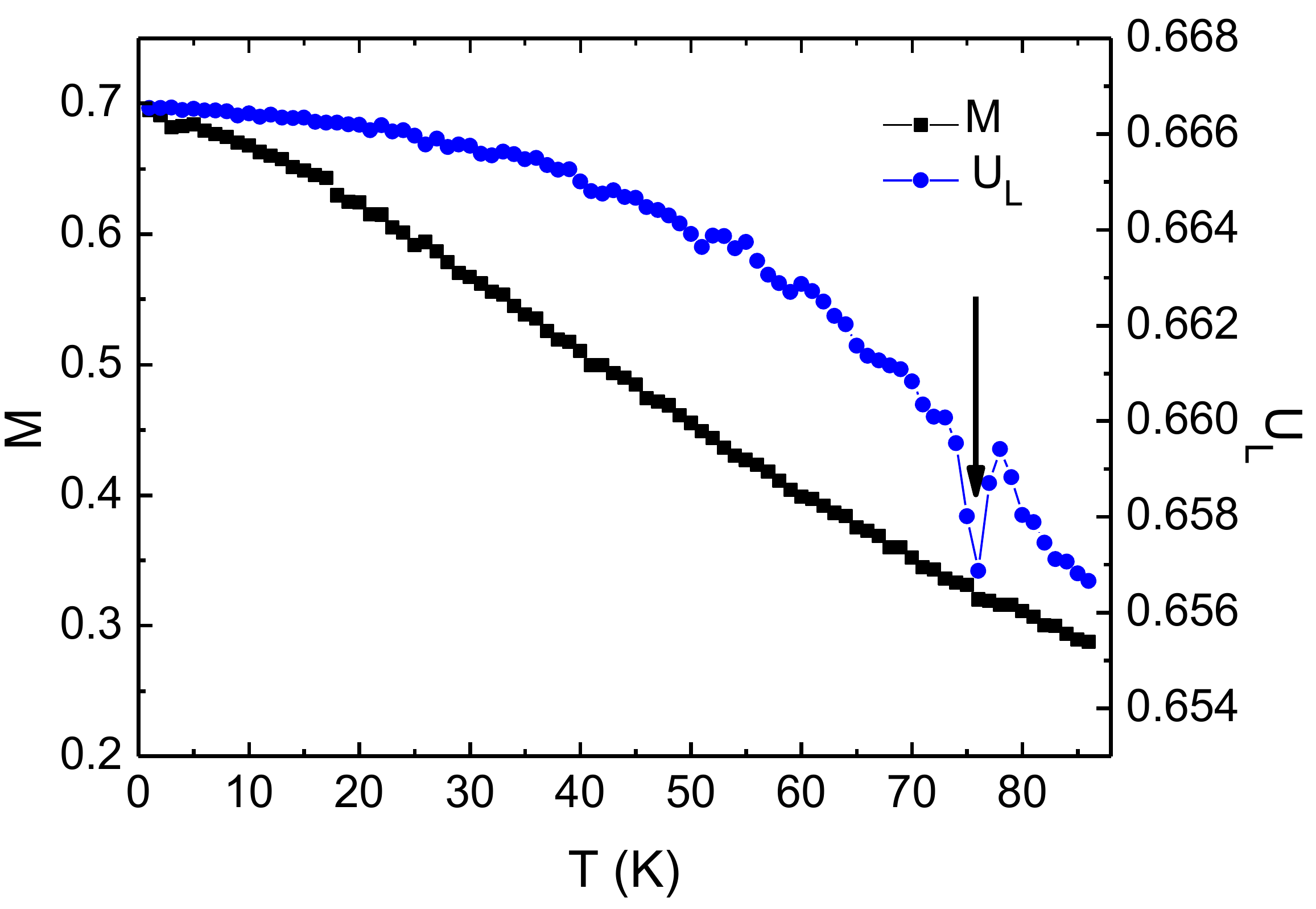}
\par\end{centering}

\caption{
Monte Carlo simulation of Bi$_{2-x}$Cr$_{x}$Se$_{3}$ with $x=0.074$. $M$ is the
simulated magnetization and $U_{L}$, Binder cumulant, is the normalized
fourth-order cumulant of the magnetization. The Curie temperature is estimated
at about 76 K.}
\label{fig:Tc}
\end{figure}

As proposed in Ref.~\onlinecite{Yu2010b}, both insulator and ferromagnetism are required to realize QAHE. After possible
candidates have been achieved, we then further investigate the feasibility of establishing ferromagnetism for Cr
and Fe in the most promising and concerned TI Bi$_{2}$Se$_{3}$, which
has the largest band gap among all the discovered TIs. First, the magnetic ground state of single magnetic dopants in
Bi$_{2}$Se$_{3}$ is identified. We have tried different initial directions of the magnetic moments. The results indicate that
both Cr and Fe prefer the direction perpendicular to the Bi$_{2}$Se$_{3}$ quintuple layers (axis 001). The magnetic moments are
about 3 $\mu_{B}$ for Cr and 5 $\mu_{B}$ for Fe, respectively. The magnetization anisotropy energies for Cr and Fe are about 7 meV
and 16 meV, respectively. Then, we investigate the magnetic coupling between the two TM dopants.
The favored magnetic state [either ferromagnetic (FM) or anti-ferromagnetic
(AFM)] is studied by calculating the total energy difference of the
two configurations at the same TM-dopant separation. Our calculations
indicate that weak AFM is favorable in Fe doped Bi$_{2}$Se$_{3}$,
while the magnetism is experimentally difficult to be observed \cite{Kulbachinskii2002,Sugama2001}
due to the scarce effective doping concentration. \cite{Zhang2012}
Our previous work \cite{Zhang2012} found that Cr doped Bi$_{2}$Se$_{3}$
prefers to be FM state, which has been also predicted by Lu \textit{et al}.~\cite{Lu2011} and confirmed by some recent experiments.
\cite{Liu2012,Haazen2012,Kou2012} The magnetic coupling strengths
[$(E_{AFM}-E_{FM})/2$] for two Cr atoms within the same QL at the
first three nearest neighboring distances are on the order of 10 meV. In our calculations, an appropriate supercell of $3\times3\times1$ Bi$_{2}$Se$_{3}$ was employed. Additional calculations from a larger supercell of $4\times4\times1$ Bi$_{2}$Se$_{3}$ indicate that magnetic coupling strengths only change within 2 meV.
From the QAHE point of view, a spontaneous FM ground state is required.
We therefore carried out Monte Carlo simulations~\cite{MonteCarloSimulationinStatisticalPhysics-AnIntroduction,Wu2007}
to determine the Curie temperature ($T_{c}$) in Cr doped Bi$_{2}$Se$_{3}$.
$L\times L\times L$ ($L$=20) Bi$_{2}$Se$_{3}$ cells with
periodic boundary conditions are used. Then the magnetic Cr atoms
are randomly distributed on the Bi lattice sites of Bi$_{2}$Se$_{3}$
with the ratio of Cr:Bi to be $x:(2-x)$, where
$x=0.074$ in the simulation. The Heisenberg Hamiltonian of the system
is described as~\cite{MonteCarloSimulationinStatisticalPhysics-AnIntroduction}

\begin{equation}
E=-\sum_{i<j}J_{ij}\overrightarrow{S_{i}}\cdot\overrightarrow{S_{j}},\label{eq:Heisenberg}
\end{equation}
 where $J_{ij}$ is the exchange coupling constant between the
$i$th and $j$th dopant atoms, taken from the first-principles calculations as shown in Table~IV.
The thermodynamic magnetization per atom can be calculated by $M(T)=<[(\sum_{i}S_{i}^{x})^{2}+(\sum_{i}S_{i}^{y})^{2}+(\sum_{i}S_{i}^{z})^{2}]^{1/2}>/N$,
where N is the number of the magnetic dopant atoms, and $<...>$ is the statistical
average over different states which are generated during the Markov process.~\cite{MonteCarloSimulationinStatisticalPhysics-AnIntroduction}
To define the Curie temperature, an accumulation of magnetization of the fourth
order $U_{L}$ (Binder-cumulant) are calculated by $U_{L}=1-<M^{4}>/3<M^{2}>^{2}$.
\cite{Binder1981and1987,Schliemann2001,Fukushima2004}
Fig.~\ref{fig:Tc} shows the simulated magnetization $M$ and $U_{L}$ as a function of
temperature for Bi$_{2-x}$Cr$_{x}$Se$_{3}$ with $x=0.074$. We get an estimated Curie
temperature $T_{c}$ at about 76 K.

\section{Summary}

In summary, we systematically studied the stability, electronic and magnetic properties
of magnetically doped topological insulators Bi$_{2}$Se$_{3}$,
Bi$_{2}$Te$_{3}$ and Sb$_{2}$Te$_{3}$ using first-principles calculations
in combination with Monte Carlo simulation. Our calculations showed that cation
site substitutional doping was energetically most favorable.
Further we suggested a recipe of effective magnetic doping for experimental
study with the optimal growth conditions. In addition, our results indicated
that under the nominal doping concentration of 4\%, Cr and Fe
doped Bi$_{2}$Se$_{3}$, Bi$_{2}$Te$_{3}$, and Cr doped Sb$_{2}$Te$_{3}$
were remain insulators, although the band gaps were substantially reduced
due to Se-dopant hybridization. Instead, all TIs doped with V and Mn as well
as Fe doped Sb$_{2}$Te$_{3}$ became metals. Finally, we explored the magnetic
coupling between dopants, suggesting FM was favorable in Cr doped Bi$_{2}$Se$_{3}$
while AFM in Fe doped material. Using Monte Carlo simulation, we estimated that the
Curie temperature of 7.4\% Cr doped Bi$_{2}$Se$_{3}$ was about 76 K.
Our results provide important guidelines towards further experimental efforts of incorporating
magnetism in TI, in particular for the realization of QAHE based on
magnetic topological insulators.

\begin{acknowledgments}

We would like to thank W.G. Zhu, D. Xiao and F. Yang for the helpful discussions.
This work was  supported by the MOST Project of China (Grants Nos. 2014CB920903, 2011CBA00100) , the NSFC (Grant Nos.
11174337, and 11225418), the Specialized Research Fund for the Doctoral Program of Higher Education of China (Grants No. 20121101110046). Z.H. was supported by NSF of China (Grants No. 11004039), National
Key Project for Basic Research of China under Grant No. 2011CBA00200.
We acknowledge support from Supercomputing Center of Chinese Academy
of Sciences.

\end{acknowledgments}


\begin{thebibliography}{100}
\bibitem{Bernevig2006} B. A. Bernevig and S.-C. Zhang, Phys. Rev.
Lett. \textbf{96}, 106802 (2006).

\bibitem{Moore2007} J. E. Moore and L. Balents, Phys. Rev. B \textbf{75},
121306 (2007).

\bibitem{Fu2007a} L. Fu, C. L. Kane, and E. J. Mele, Phys. Rev. Lett.
\textbf{98}, 106803 (2007).

\bibitem{Zhang2009} H. Zhang, C.-X. Liu, X.-L. Qi, X. Dai, Z. Fang,
and S.-C. Zhang, Nature Phys. \textbf{5}, 438 (2009).

\bibitem{Xia2009} Y. Xia, D. Qian, D. Hsieh, L. Wray, A. Pal, H.
Lin, A. Bansil, D. Grauer, Y. S. Hor, R. J. Cava, and M. Z. Hasan,
Nature Phys. \textbf{5}, 398 (2009).

\bibitem{Hsieh2009} D. Hsieh, Y. Xia, D. Qian, L. Wray, J. H. Dil,
F. Meier, J. Osterwalder, L. Patthey, J. G. Checkelsky, N. P. Ong,
A. V. Fedorov, H. Lin, A. Bansil, D. Grauer, Y. S. Hor, R. J. Cava,
and M. Z. Hasan, Nature (London) \textbf{460}, 1101 (2009).

\bibitem{Hasan2010} M. Z. Hasan and C. L. Kane, Rev. Mod. Phys. \textbf{82},
3045 (2010).

\bibitem{Qi2011_RevModPhys} X.-L. Qi and S.-C. Zhang, Rev. Mod. Phys.
\textbf{83}, 1057 (2011).

\bibitem{Chen2009} Y. L. Chen, J. G. Analytis, J.-H. Chu, Z. K. Liu,
S.-K. Mo, X. L. Qi, H. J. Zhang, D. H. Lu, X. Dai, Z. Fang, S. C.
Zhang, I. R. Fisher, Z. Hussain, and Z.-X. Shen, Science
\textbf{325}, 178 (2009).

\bibitem{Haazen2012} P. P. J. Haazen, J.-B. Lalo\"e, T. J. Nummy, H.
J. M. Swagten, P. Jarillo-Herrero, D. Heiman, and J. S. Moodera, Appl.
Phys. Lett. \textbf{100}, 082404 (2012).

\bibitem{Choi2005} J. Choi, H.-W. Lee, B.-S. Kim, S. Choi, J. Choi,
J. H. Song, and S. Cho, J. Appl. Phys. \textbf{97}, 10D324 (2005).

\bibitem{Choi2004} J. Choi, S. Choi, J. Choi, Y. Park, H.-M. Park,
H.-W. Lee, B.-C. Woo, and S. Cho, Phys. Status Solidi B \textbf{241},1541
(2004).

\bibitem{Kulbachinskii2002} V. A. Kulbachinskii, A. Y. Kaminskii,
K. Kindo, Y. Narumi, K. Suga, P. Lostak, and P. Svanda, Physica B
\textbf{311}, 292 (2002).

\bibitem{Hor2010d} Y. S. Hor, P. Roushan, H. Beidenkopf, J. Seo,
D. Qu, J. G. Checkelsky, L. A. Wray, D. Hsieh, Y. Xia, S.-Y. Xu, D.
Qian, M. Z. Hasan, N. P. Ong, A. Yazdani, and R. J. Cava, Phys. Rev.
B \textbf{81}, 195203 (2010).

\bibitem{Niu2011} C. Niu, Y. Dai, M. Guo, W. Wei, Y. Ma, and B. Huang,
Appl. Phys. Lett. \textbf{98}, 252502 (2011).

\bibitem{Dyck2002} J. S. Dyck, P. H\'{a}jek, P. Lo\v{s}t'\'{a}k, and C. Uher, Phys. Rev. B {\bf 65}, 115212 (2002); Z. Zhou, Y.-J. Chien, and C. Uher, \textit{ibid.} \textbf{74}, 224418 (2006).

\bibitem{Dyck2005} J. S. Dyck, \v{C}. Dra\v{s}ar, P. Lo\v{s}t'\'{a}k, and C. Uher, Phys. Rev. B {\bf 71}, 115214 (2005).

\bibitem{Chien2007} Y.-J. Chien, Ph.D. thesis, University of Michigan, 2007.

\bibitem{Qi2008} X.-L. Qi, T. L. Hughes, and S.-C. Zhang, Phys. Rev.
B \textbf{78}, 195424 (2008).

\bibitem{Yu2010b} R. Yu, W. Zhang, H.-J. Zhang, S.-C. Zhang, X. Dai,
and Z. Fang, Science \textbf{329}, 61 (2010).

\bibitem{Liu2009} Q. Liu, C.-X. Liu, C. Xu, X.-L. Qi, and S.-C. Zhang,
Phys. Rev. Lett. \textbf{102}, 156603 (2009).

\bibitem{Qi2009} X.-L. Qi, R. Li, J. Zang, and S.-C. Zhang, Science
\textbf{323}, 61 (2009).

\bibitem{Garate2010} I. Garate, and M. Franz, Phys. Rev. Lett. \textbf{104},
146802 (2010).

\bibitem{Wray2011b} L. A. Wray, S.-Y. Xu, Y. Xia, D. Hsieh, A. V.
Fedorov, Y. S. Hor, R. J. Cava, A. Bansil, H. Lin, and M. Z. Hasan,
Nature Phys. \textbf{7}, 32 (2011).

\bibitem{Jin2011} H. Jin, J. Im, and A. J. Freeman, Phys. Rev. B \textbf{84},
134408 (2011).

\bibitem{Zhang2013} J. Zhang, C.-Z. Chang, P. Tang, Z. Zhang,
X. Feng, K. Li, L.-L. Wang, X. Chen, C. Liu, W. Duan, K. He,
Q.-K. Xue, X. Ma, and Y. Wang, Science \textbf{339}, 1582 (2013).

\bibitem{Kim2013} H.-J. Kim, K.-S. Kim, J.-F. Wang, V. A. Kulbachinskii,
K. Ogawa, M. Sasaki, A. Ohnishi, M. Kitaura, Y.-Y. Wu, L. Li, I. Yamamoto,
J. Azuma, M. Kamada, and V. Dobrosavljevi\'{c}, Phys. Rev. Lett.
\textbf{110}, 136601 (2013).

\bibitem{Lang2013} M. Lang, L. He, X. Kou, P. Upadhyaya, Y. Fan, H. Chu,
Y. Jiang, J. H. Bardarson, W. Jiang, E. S. Choi, Y. Wang, N.-C. Yeh, J. Moore,
and K. L. Wang, Nano Lett. \textbf{13}, 48 (2013).

\bibitem{Chen2010i} Y. L. Chen, J.-H. Chu, J. G. Analytis, Z. K.
Liu, K. Igarashi, H.-H. Kuo, X. L. Qi, S. K. Mo, R. G. Moore, D. H.
Lu, M. Hashimoto, T. Sasagawa, S. C. Zhang, I. R. Fisher, Z. Hussain,
and Z. X. Shen, Science \textbf{329}, 659 (2010).


\bibitem{Henk2012} J. Henk, A. Ernst, S. V. Eremeev, E. V. Chulkov,
I. V. Maznichenko, and I. Mertig, Phys. Rev. Lett. \textbf{108},
206801 (2012). 

\bibitem{Chang2013} C.-Z. Chang, J. Zhang, X. Feng, J. Shen, Z. Zhang,
M. Guo, K. Li, Y. Ou, P. Wei, L.-L. Wang, Z.-Q. Ji, Y. Feng, S. Ji, X. Chen, J. Jia, X. Dai, Z.
Fang, S.-C. Zhang, K. He, Y. Wang, L. Lu, X.-C. Ma, and Q.-K. Xue, Science \textbf{340}, 167 (2013).

\bibitem{Zhou2006} Z. Zhou, M. \v{Z}ab\`{e}\'{i}k, P. Lo\v{s}t\'{a}k, and C. Uher,
 J. Appl. Phys. \textbf{99}, 043901 (2006).

\bibitem{He2011} H.-T. He, G. Wang, T. Zhang, I.-K. Sou, G. K. L.
Wong, J.-N. Wang, H.-Z. Lu, S.-Q. Shen, and F.-C. Zhang, Phys. Rev. Lett. \textbf{106},
166805 (2011).

\bibitem{Choi2011} Y. H. Choi, N. H. Jo, K. J. Lee, J. B. Yoon, C.
Y. You, and M. H. Jung, J. Appl. Phys. \textbf{109}, 07E312 (2011).

\bibitem{Kou2012} X. F. Kou, W. J. Jiang, M. R. Lang, F. X. Xiu,
L. He, Y. Wang, Y. Wang, X. X. Yu, A. V. Fedorov, P. Zhang, and K. L. Wang,
J. Appl. Phys. \textbf{112}, 063912 (2012).

\bibitem{Sugama2001}  Y. Sugama, T. Hayashi, H. Nakagawa, N. Miura,
and V. A. Kulbachnskii, Physica B \textbf{298}, 531 (2001).

\bibitem{Salman2012} Z. Salman, E. Pomjakushina, V. Pomjakushin,
A. Kanigel, K. Chashka, K. Conder, E. Morenzoni, T. Prokscha, K. Sedlak,
and A. Suter, arXiv:1203.4850v1.

\bibitem{Zhang1991} S. B. Zhang and J. E. Northrup, Phys. Rev. Lett.
\textbf{67}, 2339 (1991).

\bibitem{VandeWalle2004} C. G. Van de Walle and J. Neugebauer, J.
Appl. Phys. \textbf{95}, 3851 (2004).

\bibitem{Blochl1994}  P. E. Bl\"ochl, Phys. Rev. B \textbf{50}, 17953
(1994).

\bibitem{Perdew1996} J. P. Perdew, K. Burke, and M. Ernzerhof, Phys.
Rev. Lett. \textbf{77}, 3865 (1996).

\bibitem{Kresse1993a} G. Kresse and J. Hafner, Phys. Rev. B \textbf{48},
13115 (1993).

\bibitem{Kresse1996} G. Kresse and J. Furthm\"uller, Phys. Rev. B \textbf{54},
11169 (1996).

\bibitem{SOCnote} Formation energy with the inclusion of SOC is 17\%
different from non-SOC result for Bi$_{2}$Se$_{3}$.

\bibitem{West2012a} D. West, Y. Y. Sun, H. Wang, J. Bang, and S.
B. Zhang, Phys. Rev. B \textbf{86}, 121201 (2012).


\bibitem{Hor2009} Y. S. Hor, A. Richardella, P. Roushan, Y. Xia,
J. G. Checkelsky, A. Yazdani, M. Z. Hasan, N. P. Ong, and R. J. Cava,
 Phys. Rev. B \textbf{79}, 195208 (2009).

\bibitem{Wang2010} Z. Wang, T. Lin, P. Wei, X. Liu, R. Dumas, K.
Liu, and J. Shi, Appl. Phys. Lett. \textbf{97}, 042112 (2010).

\bibitem{Urazhdin2004} S. Urazhdin, D. Bilc, S. D. Mahanti, S. H. Tessmer,
T. Kyratsi, and M. G. Kanatzidis, Phys. Rev. B \textbf{69}, 085313 (2004).

\bibitem{Giani1999} A. Giani, A. Boulouz, F. Pascal-Delannoy, A.
Foucaran, E. Charles, and A. Boyer, Mater. Sci. Eng. B \textbf{64}, 19 (1999).

\bibitem{Gasenkova2001} I. V. Gasenkova, L. D. Ivanova, and Y. V.
Granatkina, Inorg. Mater. \textbf{37}, 1112 (2001).

\bibitem{Jiang2012} Y. Jiang, Y. Y. Sun, M. Chen, Y. Wang, Z. Li,
C. Song, K. He, L. Wang, X. Chen, Q.-K. Xue, X. Ma, and S. B. Zhang,
Phys. Rev. Lett. \textbf{108}, 066809 (2012).

\bibitem{Yoo2005} B. Y. Yoo, C.-K. Huang, J. R. Lim, J. Herman, M. A. Ryan,
J.-P. Fleurial, and N. V. Myung, Electrochim. Acta \textbf{50}, 4371 (2005).

\bibitem{Lee2008} J. Lee, S. Farhangfar, J. Lee, L. Cagnon, R. Scholz,
U. G\"osele, and K. Nielsch, Nanotechnology \textbf{19}, 365701 (2008).

\bibitem{Wang2012-arXiv} S.-X. Wang, P. Zhang, and S.-S. Li, arXiv:1201.2469

\bibitem{Hashibon2011} A. Hashibon, and C. Els\"asser,
 Phys. Rev. B \textbf{84}, 144117 (2011).

\bibitem{Wang2011b} G. Wang, X.-G. Zhu, Y.-Y. Sun, Y.-Y. Li, T. Zhang,
J. Wen, X. Chen, K. He, L.-L. Wang, X.-C. Ma, J.-F. Jia, S. B. Zhang,
and Q.-K. Xue, Adv. Mater. \textbf{23}, 2929 (2011).


\bibitem{Wang2011a} Y.-L. Wang, Y. Xu, Y.-P. Jiang, J.-W. Liu, C.-Z. Chang, M. Chen, Z. Li,
C.-L. Song, L.-L. Wang, K. He, X. Chen, W.-H. Duan, Q.-K. Xue, X.-C. Ma,
 Phys. Rev. B \textbf{84}, 075335 (2011).

\bibitem{West2012} D. West, Y. Y. Sun, S. B. Zhang, T. Zhang, X.
Ma, P. Cheng, Y. Y. Zhang, X. Chen, J. F. Jia, and Q. K. Xue,
 Phys. Rev. B \textbf{85}, 081305 (2012).

\bibitem{Song2012} C.-L. Song, Y.-P. Jiang, Y.-L. Wang, Z. Li, L.
Wang, K. He, X. Chen, X.-C. Ma, and Q.-K. Xue,
 Phys. Rev. B \textbf{86}, 045441 (2012).

\bibitem{Zhang2012} J.-M. Zhang, W. Zhu, Y. Zhang, D. Xiao, and Y.
Yao, Phys. Rev. Lett. \textbf{109}, 266405 (2012).

\bibitem{Liu2012} M. Liu, J. Zhang, C.-Z. Chang, Z. Zhang, X. Feng,
K. Li, K. He, L.-l. Wang, X. Chen, X. Dai, Z. Fang, Q.-K. Xue, X.
Ma, and Y. Wang, Phys. Rev. Lett. \textbf{108}, 036805 (2012).

\bibitem{Cha2010}  J. J. Cha, J. R. Williams, D. Kong, S. Meister,
H. Peng, A. J. Bestwick, P. Gallagher, D. Goldhaber-Gordon, and Y.
Cui, Nano Lett. \textbf{10}, 1076 (2010).

\bibitem{Song2010} C.-L. Song, Y.-L. Wang, Y.-P. Jiang, Y. Zhang,
C.-Z. Chang, L. Wang, K. He, X. Chen, J.-F. Jia, Y. Wang, Z. Fang,
X. Dai, X.-C. Xie, X.-L. Qi, S.-C. Zhang, Q.-K. Xue, and X. Ma, Appl.
Phys. Lett. \textbf{97}, 143118 (2010).


\bibitem{a-Larson2008} P. Larson, and W. R. L. Lambrecht,
Phys. Rev. B \textbf{78}, 195207 (2008).

\bibitem{Zhang2010} W. Zhang, R. Yu, H.-J. Zhang, X. Dai, and Z. Fang,
New J. Phys. \textbf{12}, 065013 (2010). 

\bibitem{Lu2011} H.-Z. Lu, J. Shi, and S.-Q. Shen, Phys. Rev. Lett.
\textbf{107}, 076801 (2011).

\bibitem{MonteCarloSimulationinStatisticalPhysics-AnIntroduction}
K. Binder and D. Heermann, Monte Carlo Simulation in Statistical Physics:
An Introduction (Springer, Berlin, 2010), 5th ed.

\bibitem{Wu2007} Q. Wu, Z. Chen, R. Wu, G. Xu, Z. Huang, F. Zhang,
and Y. Du, Solid State Commun. \textbf{142}, 242 (2007).

\bibitem{Binder1981and1987} K. Binder, Z. Phys. B \textbf{43}, 119 (1981);
K. Binder, Rep. Prog. Phys. \textbf{50}, 783 (1987).

\bibitem{Schliemann2001} J. Schliemann, J. K\"onig, and A. H. MacDonald,
Phys. Rev. B \textbf{64}, 165201 (2001).

\bibitem{Fukushima2004} T. Fukushima, K. Sato, H. Katayama-Yoshida,
and P. H. Dederichs, Jpn. J. Appl. Phys. \textbf{43}, L1416 (2004).

\end{thebibliography}
\end{document}